# scientific reports

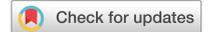

## OPEN  Generalized monodromy method in gauge/gravity duality

Yuanpeng Hou

The method of monodromy is an important tool for computing Virasoro conformal blocks in a two-dimensional Conformal Field Theory (2d CFT) at large central charge and external dimensions. In deriving the form of the monodromy problem, which defines the method, one needs to insert a degenerate operator, usually a level-two operator, into the corresponding correlation function. It can be observed that the choice of which degenerate operator to insert is arbitrary, and they shall reveal the same physical principles underlying the method. In this paper, we exploit this freedom and generalize the method of monodromy by inserting higher-level degenerate operators. We illustrate the case with a level-three operator, and derive the corresponding form of the monodromy problem. We solve the monodromy problem perturbatively and numerically; and check that it agrees with the standard monodromy method, despite the fact that the two versions of the monodromy problem do not seem to be related in any obvious way. The forms corresponding to other higher-level degenerate operators are also discussed. We explain the physical origin of the coincidence and discuss its implication from a mathematical perspective.

The AdS/CFT correspondence, also known as the gauge/gravity duality, is considered as an explicit realization of the holographic principle[1–3]. It claims that a weakly coupled quantum gravity theory in $d+1$-dimensional asymptotic AdS spacetime is dual to a $d$-dimensional strongly coupled gauge field theory on the boundary of AdS spacetime. The first concrete realization of AdS/CFT correspondence was proposed in 1997 by Maldacena in the context of string theory: Type IIB string on $AdS_5 \times S^5$ is dual to $\mathcal{N} = 4\, U(N)$ Super Yang-Mills theory[4]. It was then followed by more realizations such as the $AdS_4/CFT_3$ version between ABJM theory and M theory[5]. By now there have accumulated reasonable grounds to suppose that the duality should hold in a more general context, even though it has not been proved apart from a few examples mentioned before, particularly due to the lack of understanding towards a full theory of quantum gravity. The AdS/CFT correspondence has been applied to gain novel understandings of many interesting physics, such as the study of black hole thermal dynamics[6], non-Fermi liquids and quantum phase transitions[7], holographic entanglement entropy[8,9], energy conditions in quantum gravity[10–12], etc., and continues to inspire new ideas in various fields.

Among the various dimensions in which AdS/CFT has been formulated, $AdS_3/CFT_2$ is special due to additional simplifications[13–16]. In particular, the symmetry of a two-dimensional conformal field theory is enhanced to the infinite-dimensional Virasoro algebra, while the bulk gravitational dynamics is topological when bulk matters are not considered[15,17–20]. These features simplify the analysis significantly, and many computations can be performed with better control. In this correspondence, the central charge $c$ of the Conformal Field Theory (CFT) is related to scale of AdS radius $l$ in Newton's constant $G$ by[16]:

$$c = \frac{3l}{2G} \qquad (1)$$

So the semi-classical limit of bulk gravity corresponds to the large $c$ limit of CFT. The asymptotic symmetry of quantum gravity in $AdS_3$ is exactly the Virasoro algebra in 2d CFT[16], in this sense and the energy–momentum tensor $T$ in 2d CFT acting on the vacuum will create bulk states containing "Boundary Graviton" in $AdS_3$. Contributions from these gravitons can be organized into Virasoro families, which are representations of Virasoro generators $\{L_{-n}, n > 0\}$, in this way we have a better understanding in the boundary CFT of how to isolate the dynamics of bulk gravitons in $AdS_3$[21–23].

Correlation functions play central roles in 2d CFT, they can reveal many important aspects of the theory. While the two-point and three-point functions are universally fixed by the symmetries of the CFTs, starting from four-point functions they contain richer information of the underlying CFT data such as the operator spectrum.

[1]Xinjiang Astronomical Observatory, Chinese Academy of Sciences, 150 Science 1-Street, Urumqi 830011, China. [2]Kavli Institute for Theoretical Sciences (KITS), University of Chinese Academy of Sciences, Beijing 100190, China. [3]School of Astronomy and Space Science, University of Chinese Academy of Sciences, Beijing 100049, China. email: houyuanpeng@xao.ac.cn

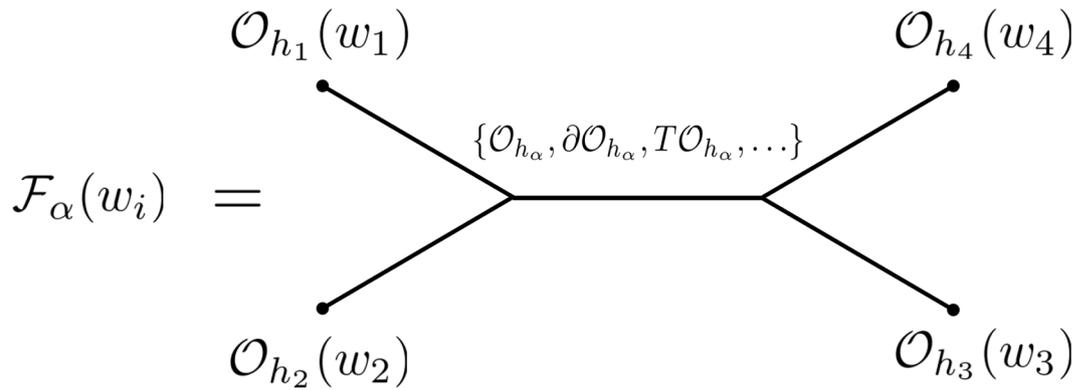

**Figure 1.** Virasoro conformal block in diagrammatic language.

Using the infinite-dimensional Virasoro algebra of the underlying symmetry, the four-point correlation function computations can be approached by decomposing them into Virasoro conformal blocks, which include the Operator Product Expansion (OPE) contributions from all operators created from a particular primary operator by all combinations of Virasoro creation generators $\{L_{-n}, n > 0\}$. This is to be contrasted with the usual decomposition into global conformal blocks as in higher dimensional CFTs, which include only contributions from the global generators $\{L_{-1}, L_{-0}, L_1\}$. In practice, the Virasoro conformal blocks are labeled by the conformal dimensions of the external operators $\{h_i, i = 1, \ldots, 4\}$, the central charge $c$, and the conformal dimension $h_\alpha$ of the internal operator being exchanged that defines the block, see Fig. 1.

Unlike the global blocks whose analytic forms are known to be hypergeometric functions, the full analytic forms of the Virasoro conformal blocks remain unknown. As a result, various methods have been developed to compute the explicit expressions for Virasoro conformal blocks in various limits, such as the Conformal Casimir approach[24], Zamolodchikov's recursion relation[25–27], the method of monodromy[21,28–31], etc.

In this paper we focus on the method of monodromy. This technique is particularly useful for computing Virasoro conformal blocks in the semi-classical limit, which is the limit defined by sending $c \to \infty$ while holding $\{h_i/c, i = 1, \ldots, 4\}$ fixed. We shall review the derivation of the method in the next section, but the main idea is as follows. By inserting a level-two degenerate operator $\hat{\psi}_2$ into the correlation function and invoking the degeneracy condition, we obtain a second order differential equation for the five-point function $\left\langle O_1 O_2 \hat{\psi}_2(z) O_3 O_4 \right\rangle$. If we further assume that the five-point function can be factorized into the product of a "Wave Function" $\psi_2(z)$ multiplying the original four-point function, which is expected to hold in the semi-classical limit:

$$\left\langle O_1 O_2 \hat{\psi}_2(z) O_3 O_4 \right\rangle \approx \psi_2(z) \langle O_1 O_2 O_3 O_4 \rangle \tag{2}$$

We end up with a differential equation for the "Wave Function" $\psi_2(z)$ that also depends on the value of the four-point function. By imposing that the solutions satisfy certain monodromy condition which depends on $h_\alpha$, we end up solving the contribution to the four-point function from a particular Virasoro conformal block associated with the primary operator $\alpha$. The monodromy method can be derived explicitly in Liouville Field Theory (LFT)[28]. However, since the Virasoro conformal blocks are fixed only by the Virasoro algebra, agnostic to the underlying theory, the method should be applicable to any CFT. The monodromy method has been developed and applied to various situations. For example, generalizing the method for the computation of conformal blocks to the torus is developed in[32]. Higher-point conformal blocks and the entanglement entropy of an arbitrary number of disjoint intervals for heavy states have been calculated in[33].

By observing the derivation of the standard monodromy method carefully, it is easily recognized that the derivation involves making choices that reflect certain "Moving Components" of the method. In particular, the decision for inserting a level-two degenerate operator $\hat{\psi}_2$ is somehow arbitrary, not dictated by any necessary condition. While practically this choice possibly gives rise to the simplest realization of the method—involving differential equations of the lowest order, it is interesting to explore other possibilities. As one may expect, different choices of degenerate operators mean different degeneracy conditions, which will generate different versions of the monodromy problem. However, these monodromy problems are not at all un-related: solving them gives rise to the same Virasoro conformal block. While this connection is obvious from a "physics" point of view—they reflect the validity of ansatz (2) for inserting different degenerate operators in the semi-classical limit, they do not appear to be obvious from a mathematical point of view. In this paper, the main goal is to explore and examine such connections, and to discuss the implication of such connections from a mathematical perspective.

The paper is organized as follows. In "Standard monodromy method" section, we give a review of the derivation for the standard monodromy method. In "Level-three monodromy method" section, we generalize the derivation to the particular case of inserting a level-three degenerate operator into the correlation function, and obtain the corresponding form of the monodromy method. In "Computing vacuum block using level-three monodromy method" section, we compare the two monodromy methods by first computing the four-point vacuum Virasoro blocks in the heavy-light limit perturbatively, and then checking it numerically away from the perturbative limit. The perfect agreements between the results verify the consistency between the methods and thus confirm the connection. In "Generalizing to higher-level monodromy method" section, we further generalize our method to higher-level degenerate operators. We obtain a very general version of the monodromy method and discuss its application. Finally, we conclude with a discussion about the connection between different monodromy methods and an outlook for the methods.

## Standard monodromy method

### Review of Virasoro conformal block decomposition.

We first give a brief review of the standard monodromy method for computing Virasoro conformal blocks. Our review of the method mainly follows[21,29–31]. We consider the four-point correlation function in 2d CFT and decompose it into conformal blocks:

$$\begin{aligned}\langle O_1(w_1)O_2(w_2)O_3(w_3)O_4(w_4)\rangle &= \sum_p \langle O_1(w_1)O_2(w_2)|p\rangle\langle p|O_3(w_3)O_4(w_4)\rangle \\ &= \sum_{\alpha,n} \langle O_1(w_1)O_2(w_2)|\alpha;n\rangle\langle \alpha;n|O_3(w_3)O_4(w_4)\rangle \\ &= \sum_\alpha F_\alpha(w_i)\end{aligned} \quad (3)$$

where $O_i$ for $i = 1, \ldots, 4$ are primary operators with conformal dimensions $h_i$, which are inserted at points $w_i$; $|p\rangle$ denotes all possible intermediate states of the theory. In the second line, we organized the states $|p\rangle$ into Virasoro families $|\alpha;n\rangle$ where α denotes the conformal block associated with primary operator α with conformal dimension $h_\alpha$. $F_\alpha(w_i)$ captures the contribution from all Virasoro descendants of α.

An explicit expression for $F_\alpha(w_i)$ is not known in general. For generic $c$ and $\{h_i, h_\alpha\}$ it admits a series expansion in conformal ratio $x = \frac{(w_1-w_2)(w_3-w_4)}{(w_1-w_3)(w_2-w_4)}$ using Zamolodchikov's recursion relation[25–27]. For generic $x$, the method of monodromy provides a complementary tool for computing the conformal block in the semi-classical limit, which as mentioned before corresponds to the limit of sending $c \to \infty$ while holding the ratios $h_i/c$ finite. In this limit, we can approximate the conformal blocks as:

$$F_\alpha(w_i) \approx e^{-\frac{c}{6}f_\alpha(w_i)} \quad (4)$$

The ansatz (4) was conjectured in[34] and has been proven in[35] recently. For later purposes we define the so-called accessory parameters, denoted by $c_i$:

$$c_i = \partial_{w_i} f_\alpha(w_i) \quad (5)$$

The monodromy method works by fixing the accessory parameters $c_i$ so that we can integrate them to get the conformal blocks $F_\alpha(w_i)$ via (4) and (5).

### From degenerate operator to differential equation.

To derive the monodromy method, one starts by inserting a degenerate operator $\hat{\psi}_2$ at level two into the correlation function. The degeneracy condition corresponds to requiring that via state-operator correspondence it creates a primary state $|\psi_2\rangle$ and the following descendant:

$$|\chi_{2,1}\rangle = [L_{-2} + \zeta L_{-1}^2]|\psi_2\rangle \quad (6)$$

For some ζ is null, i.e. $\langle\chi_{2,1}|\chi_{2,1}\rangle = 0$. From this one can solve the corresponding conformal dimension $h_{\psi_2}$ of $|\psi_2\rangle$ and coefficient ζ:

$$\begin{cases}\zeta = -\frac{3}{2(2h_{\psi_2}+1)} \\ h_{\psi_2} = \frac{5-c-\sqrt{(c-1)(c-25)}}{16}\end{cases} \quad (7)$$

It is conventional to adopt the reparameterization at large $c$:

$$c = 1 + 6\left(b + \frac{1}{b}\right)^2 \overset{b \ll 1}{\sim} \frac{6}{b^2} \quad (8)$$

And then we get:

$$\begin{cases}\zeta \approx \frac{c}{6} \\ h_{\psi_2} \approx -\frac{1}{2} - \frac{3b^2}{4}\end{cases} \quad (9)$$

Now, we insert the degenerate operator into the four-point correlation function and use the degeneracy condition $|\chi_{2,1}\rangle = 0$:

$$\begin{aligned}\langle O_1 O_2 \hat{\chi}_{2,1}(z) O_3 O_4\rangle &= \langle O_1 O_2 (L_{-2} + \zeta L_{-1}^2)\hat{\psi}_2(z) O_3 O_4\rangle \\ &= (L_{-2} + \zeta L_{-1}^2)\langle O_1 O_2 \hat{\psi}_2(z) O_3 O_4\rangle \\ &= 0\end{aligned} \quad (10)$$

By the formula:

$$L_{-1} = \partial_z, \quad L_{-m} = \sum_i \left( \frac{(m-1)h_i}{(w_i - z)^m} - \frac{1}{(w_i - z)^{m-1}} \partial_{w_i} \right), \quad m \geq 2 \tag{11}$$

We have:

$$\left( \sum_{i=1}^{4} \left( \frac{h_i}{(z - w_i)^2} + \frac{1}{z - w_i} \partial_{w_i} \right) + \zeta \partial_z^2 \right) \langle O_1 O_2 \hat{\psi}_2(z) O_3 O_4 \rangle = 0 \tag{12}$$

Decomposing the correlation function into conformal blocks, we get:

$$\left( \frac{c}{6} \partial_z^2 + \sum_{i=1}^{4} \left( \frac{h_i}{(z - w_i)^2} + \frac{1}{z - w_i} \partial_{w_i} \right) \right) \langle O_1 O_2 \hat{\psi}_2(z) O_3 O_4 \rangle$$

$$= \sum_p \left( \frac{c}{6} \partial_z^2 + \sum_{i=1}^{4} \left( \frac{h_i}{(z - w_i)^2} + \frac{1}{z - w_i} \partial_{w_i} \right) \right) \langle O_1 O_2 | p \rangle \langle p | \hat{\psi}_2(z) O_3 O_4 \rangle \tag{13}$$

$$= \sum_{\alpha,n} \left( \frac{c}{6} \partial_z^2 + \sum_{i=1}^{4} \left( \frac{h_i}{(z - w_i)^2} + \frac{1}{z - w_i} \partial_{w_i} \right) \right) \psi_2(z) \langle O_1 O_2 | \alpha; n \rangle \langle \alpha; n | O_3 O_4 \rangle$$

The third line of (13) uses the ansatz (2)—the five-point function $\langle O_1 O_2 \hat{\psi}_2(z) O_3 O_4 \rangle$ can be factorized into the product of a "Wave Function" $\psi_2(z)$ multiplying the original four-point function $\langle O_1 O_2 O_3 O_4 \rangle$, which is expected to hold in the semi-classical limit. The key ansatz encodes the main physical picture behind the validity of the monodromy method. Using (3) and exponential form of the conformal block $F_\alpha(w_i) \approx e^{-\frac{c}{6} f_\alpha(w_i)}$, we get:

$$\sum_\alpha \left( \frac{c}{6} \partial_z^2 + \sum_{i=1}^{4} \left( \frac{h_i}{(z - w_i)^2} + \frac{1}{z - w_i} \partial_{w_i} \right) \right) \psi_2(z) F_\alpha(w_i)$$

$$\approx \sum_\alpha \left( \frac{c}{6} \partial_z^2 + \sum_{i=1}^{4} \left( \frac{h_i}{(z - w_i)^2} + \frac{1}{z - w_i} \partial_{w_i} \right) \right) \psi_2(z) e^{-\frac{c}{6} f_\alpha(w_i)} \tag{14}$$

$$= 0$$

As we shall see, distinct conformal blocks $\alpha$ need to satisfy monodromy condition that depends on $h_\alpha$, which are independent of one another. As a result, each conformal block should satisfy the differential equation separately. So for each conformal block:

$$\left( \frac{c}{6} \partial_z^2 + \sum_{i=1}^{4} \left( \frac{h_i}{(z - w_i)^2} + \frac{1}{z - w_i} \partial_{w_i} \right) \right) \psi_2(z) e^{-\frac{c}{6} f_\alpha(w_i)} = 0 \tag{15}$$

We can neglect $\partial_{w_i}$ derivatives acting on $\psi_2(z)$ since $\psi_2(z)$ scales like $O(c^0)$, $\partial_{w_i} e^{-\frac{c}{6} f_\alpha(w_i)}$ is the dominant term in $\partial_{w_i} \left( \psi_2(z) e^{-\frac{c}{6} f_\alpha(w_i)} \right)$, then we have:

$$\frac{c}{6} \left( \psi_2''(z) + \sum_{i=1}^{4} \left( \frac{6h_i/c}{(z - w_i)^2} - \frac{f_\alpha'(w_i)}{z - w_i} \right) \psi_2(z) \right) e^{-\frac{c}{6} f_\alpha(w_i)} = 0 \tag{16}$$

So,

$$\psi_2''(z) + \sum_{i=1}^{4} \left( \frac{\varepsilon_i}{(z - w_i)^2} - \frac{c_i}{z - w_i} \right) \psi_2(z) = 0 \tag{17}$$

Then we get the equation:

$$\psi_2''(z) + T(z, w_i) \psi_2(z) = 0 \tag{18}$$

where we have defined:

$$T(z, w_i) = \sum_{i=1}^{4} \left( \frac{\varepsilon}{(z - w_i)^2} - \frac{c_i}{z - w_i} \right), \quad \varepsilon_i \equiv \frac{6h_i}{c} \tag{19}$$

So we get the differential equation of the standard monodromy method (18), which is also known as the decoupling equation. This equation belongs to the Heun Eq. (108), whose basics we briefly summarize in the Appendix. Three of the accessory parameters $c_i$ can be fixed by the regularity of $T(z, w_i)$ at $z = \infty$, due to the absence of operator insertion there. As a result, $T(z, w_i)$ must decay like $\frac{1}{z^4}$ as $z \to \infty$, which amounts to the conditions:

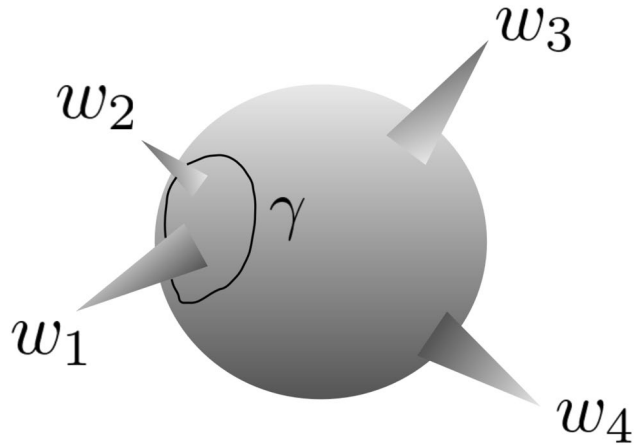

**Figure 2.** Differential equation of the standard monodromy method has regular singularities at points $z=w_i$ for $i=1,\ldots,4$ on the Riemann Sphere. "Wave Function" $\psi_2(z)$ has a monodromy if we take $z$ go through a contour $\gamma$ that encircles $w_1$ and $w_2$.

$$\sum_{i=1}^{4} c_i = 0, \quad \sum_{i=1}^{4}\left(c_i w_i - \frac{6h_i}{c}\right) = 0, \quad \sum_{i=1}^{4}\left(c_i w_i^2 - \frac{12h_i}{c}w_i\right) = 0 \quad (20)$$

Now, we apply a global conformal transformation sending $(w_1, w_2, w_3, w_4)$ to $(0, x, 1, \infty)$, where $x$ is conformal ratio $\frac{(w_1-w_2)(w_3-w_4)}{(w_1-w_3)(w_2-w_4)}$. The accessory parameters transform accordingly:

$$\begin{cases} c_1 = (x-1)c_2 - \frac{6(h_1+h_2+h_3-h_4)}{c} \\ c_3 = -xc_2 + \frac{6(h_1+h_2+h_3-h_4)}{c} \\ c_4 = 0 \end{cases} \quad (21)$$

Plugging back into (19), we will obtain the final form of $T(z, x)$ and the differential equation:

$$\begin{cases} \psi_2''(z) + T(z,x)\psi_2(z) = 0 \\ T(z,x) = \frac{\varepsilon_1}{z^2} + \frac{\varepsilon_2}{(z-x)^2} + \frac{\varepsilon_3}{(z-1)^2} + \frac{\varepsilon_1+\varepsilon_2+\varepsilon_3-\varepsilon_4}{z(1-z)} - \frac{c_2(x)(1-x)x}{z(z-x)(1-z)} \end{cases} \quad (22)$$

Notice that now the differential equation depends only on a single accessory parameter $c_2(x)$.

**Monodromy around singular points.** In order to determine $c_2(x)$, from which the conformal blocks $F_\alpha(x)$ can be obtained, we need to use the fact that $\psi(z)$ (For conciseness, we have omitted the subscript "2" in $\psi_2(z)$) has a monodromy if we encircle singularities $w_1$ and $w_2$. Figure 2 shows the points $z=w_i$ on the Riemann sphere for the differential equation of the standard single value method.

Let us consider $\psi(z)$:

$$\begin{aligned}\psi(z) &= \frac{\langle O_1 O_2 \hat{\psi}(z) O_3 O_4 \rangle}{\langle O_1 O_2 O_3 O_4 \rangle} \\ &= \frac{\sum_{\alpha,n}\langle O_1 O_2|\alpha;n\rangle\langle\alpha;n|\hat{\psi}(z)O_3 O_4\rangle}{\sum_{\alpha,n}\langle O_1 O_2|\alpha;n\rangle\langle\alpha;n|O_3 O_4\rangle} \\ &\approx \frac{\langle\alpha|\hat{\psi}(z)O_3 O_4\rangle}{\langle\alpha|O_3 O_4\rangle} \\ &= \frac{\langle 0|O_\alpha(y)\hat{\psi}(z)O_3 O_4|0\rangle}{\langle\alpha|O_3 O_4\rangle}\end{aligned} \quad (23)$$

The key step is in the third line and the approximation has been argued in[21]. The main content of the argument is as follows. We define the mode $\psi_n(z) \equiv \frac{\langle O_1 O_2|\alpha;n\rangle\langle\alpha;n|\hat{\psi}(z)O_3 O_4\rangle}{\langle O_1 O_2|\alpha;n\rangle\langle\alpha;n|O_3 O_4\rangle}$ first and we can argue the relation $\psi_{n+l}(z) \approx \psi_n(z) \sim O(c^0)$, where $l$ is the increased level. Then inserting the relation into the second line of (23) will lead to the third line. We take the leading order of the OPE: $O_\alpha(y)\hat{\psi}(z) \sim (z-y)^k O_d(y)$ as $z \to y$ in the last line of (23), so we have:

$$\psi(z) = (z - y)^k \frac{\langle 0|O_d(y)O_3O_4|0\rangle}{\langle \alpha|O_3O_4\rangle} \sim (z - y)^k \tag{24}$$

where $k$ can be determined by inserting this formula into the differential equation of correlation function (12):

$$\left( \frac{1}{b^2} \partial_z^2 + \frac{h_\alpha}{(z-y)^2} + \frac{1}{z-y} \partial_y + \sum_{i=3,4} \left( \frac{h_i}{(z-w_i)^2} + \frac{1}{z-w_i} \partial_{w_i} \right) \right) (z-y)^k \langle O_2 O_3 O_4 \rangle = 0 \tag{25}$$

Then we use the limits $c \to \infty, z \to y$, so we have $b^2 \to 0, |z - w_i| \gg |z - y|$. It is straightforward to get:

$$k^2 - (1 + b^2)k + h_\alpha b^2 \approx 0 \tag{26}$$

We can solve this equation and immediately get:

$$k \approx \frac{1 \pm \sqrt{1 - 4h_\alpha b^2}}{2} \tag{27}$$

In fact, it reproduces the information regarding the fusion rule of the level-two degenerate operator $\hat{\psi}$ with $O_\alpha$: $\hat{\psi} \times O_\alpha = O_{\beta_1} + O_{\beta_2}$, where the conformal dimesions of $O_{\beta_1}$ and $O_{\beta_2}$ are $h_\alpha - \frac{\sqrt{1-c}+\sqrt{25-c}}{\sqrt{24}}$ and $h_\alpha + \frac{\sqrt{1-c}+\sqrt{25-c}}{\sqrt{24}}$, respectively. By (27), we have:

$$\psi(z) \sim (z - y)^{\frac{1 \pm \Lambda_\alpha}{2}}, \quad \Lambda_\alpha \equiv \sqrt{1 - 4h_\alpha b^2} \tag{28}$$

Let us calculate the monodromy of the first order solution around the point $y$. We take $z$ go through a cycle that encircles point $y$. Because the OPE limit $\{w_1, w_2\} \to y$, this amounts to a contour that also encloses both $w_1$ and $w_2$. Then $\hat{\psi}$ should behave as follows:

$$\begin{aligned} \psi_1(z) &\to e^{i\pi(1+\Lambda_\alpha)} \psi_1(z) \\ \psi_2(z) &\to e^{i\pi(1-\Lambda_\alpha)} \psi_2(z) \end{aligned} \tag{29}$$

where subscripts 1 and 2 stand for two different solutions. We write this relation by matrix form:

$$\begin{pmatrix} \psi_1 \\ \psi_2 \end{pmatrix} \to \begin{pmatrix} e^{i\pi(1+\Lambda_\alpha)} & 0 \\ 0 & e^{i\pi(1-\Lambda_\alpha)} \end{pmatrix} \begin{pmatrix} \psi_1 \\ \psi_2 \end{pmatrix} \tag{30}$$

We define the monodromy matrix as:

$$M_2 \equiv \begin{pmatrix} e^{i\pi(1+\Lambda_\alpha)} & 0 \\ 0 & e^{i\pi(1-\Lambda_\alpha)} \end{pmatrix} = -\begin{pmatrix} e^{i\pi \Lambda_\alpha} & 0 \\ 0 & e^{-i\pi \Lambda_\alpha} \end{pmatrix} \tag{31}$$

So we have:

$$\begin{pmatrix} \psi_1 \\ \psi_2 \end{pmatrix} \to M_2 \begin{pmatrix} \psi_1 \\ \psi_2 \end{pmatrix} \tag{32}$$

and the trace of the monodromy equation, which doesn't depend on the basis in which the two solutions of Eq. (3) are decomposed, is:

$$Tr M_2 = -2 \cos(\pi \Lambda_\alpha) \tag{33}$$

For example, if we want to calculate the identity (vacuum) conformal block with $h_\alpha = 0$, the corresponding $M_2$ is a $2 \times 2$ identity matrix and its trace $Tr M_2 = 2$. The monodromy matrix and its trace therefore depend on $\varepsilon_i$, $x$, and $c_2$. One can thus view (33) as an equation that one can use to solve $c_2$ as a function of $\varepsilon_i$ and $x$. We call this the monodromy equation. It is in general some complicated transcendental equation that is difficult to solve analytically. There exist various limits in which the monodromy equation can be solved explicitly. For example, there is a limit called the heavy-light limit, in which we take $\varepsilon_3 = \varepsilon_4 = \varepsilon_H \gg 1$ to be heavy, and treat $\varepsilon_1 = \varepsilon_2 = \varepsilon_L \ll 1$ to be a small parameter, and one can solve the monodromy equation perturbatively in $\varepsilon_L$. The detailed steps of using the standard level-two monodromy method to compute the four-point Virasoro conformal blocks perturbatively in the semi-classical limit can be found in[21,29]. The procedure computes the conformal block in a particular channel, i.e. $(12 \to \alpha \to 34)$; in order to calculate it in a different channel, say $(13 \to \alpha \to 24)$, one needs to impose the monodromy condition around a circle enclosing $w_1$ and $w_3$.

## Level-three monodromy method
**Third order differential equation.** As has been alluded to in the introduction, the form of the monodromy problem as derived in "Standard monodromy method" section. is dictated by the choice of inserting a level-two degenerate operator $\hat{\psi}_2$ into the correlation function. At the technical level, this is a free choice. We could have inserted other degenerate operator into the correlation function. As long as the degenerate operator does not become too heavy, i.e. its conformal dimension does not scale with $c$, the approximations in "Standard monodromy method" section. still carry through. As a result, some variants of the standard monodromy method should emerge at the end of the analysis. It is interesting to explore the forms and implications of these variants. In this section, we initiate this investigation by explicitly deriving a monodromy method that stems from inserting a level-three operator $\hat{\psi}_3$ into the correlator. The derivation of this monodromy method proceeds in an analogous way to the standard monodromy method. First, we give the corresponding null state and the conformal dimension associated with $\hat{\psi}_3$:

$$|\chi_{3,1}\rangle = \left[L_{-3} - \frac{2}{2+h_{\psi_3}}L_{-1}L_{-2} + \frac{1}{(1+h_{\psi_3})(2+h_{\psi_3})}L_{-1}^3\right]|\psi_3\rangle$$
$$h_{\psi_3} = \frac{1}{6}(7 - c + \sqrt{(c-1)(c-25)})$$
(34)

For conciseness, we adopt reparameterization again $\xi \equiv -\frac{2}{h_{\psi_3}+2}$, $\eta \equiv \frac{1}{(h_{\psi_3}+1)(h_{\psi_3}+2)}$, $c = 1 + 6(b + \frac{1}{b})^2 \overset{b \ll 1}{\sim} \frac{6}{b^2}$. So we have:

$$|\chi_{3,1}\rangle = [L_{-3} + \xi L_{-1}L_{-2} + \eta L_{-1}^3]|\psi_3\rangle$$
$$h_{\psi_3} = -\frac{3b^2}{4} - \frac{3}{8}$$
(35)

We now insert the corresponding degenerate condition for the operator $\hat{\psi}_3$ into the four-point correlation function:

$$\begin{aligned}\langle O_1 O_2 \hat{\chi}_3(z) O_3 O_4 \rangle &= \langle O_1 O_2 \left(L_{-3} + \xi L_{-1}L_{-2} + \eta L_{-1}^3\right)\hat{\psi}_3(z) O_3 O_4 \rangle \\
&= \left(L_{-3} + \xi L_{-1}L_{-2} + \eta L_{-1}^3\right)\langle O_1 O_2 \hat{\psi}_3(z) O_3 O_4 \rangle \\
&= \left(\sum_{i=1}^4 \left(\frac{2h_i}{(z-w_i)^3} - \frac{1}{(z-w_i)^2}\partial_{w_i}\right) + \xi \partial_z \sum_{i=1}^4 \left(\frac{h_i}{(z-w_i)^2} - \frac{1}{(z-w_i)}\partial_{w_i}\right) + \eta \partial_z^3\right)\langle O_1 O_2 \hat{\psi}_3(z) O_3 O_4 \rangle \\
&= 0\end{aligned}$$
(36)

Just like we did in the standard monodromy method, we use the ansatz (2) again. So we can factorize the five-point correlation function into "Wave Function" $\hat{\psi}_3$ multiplying the original four-point function in the semi-classical limit. It ensures that the different "Wave Function" $\hat{\psi}_2$ and $\hat{\psi}_3$ can multiply the same four-point correlation function $\langle O_1 O_2 O_3 O_4 \rangle$ and then leads to the important differential equations. Then we have:

$$\begin{aligned}\langle O_1 O_2 \hat{\psi}_3(z) O_3 O_4 \rangle &= \psi_3(z) \langle O_1 O_2 O_3 O_4 \rangle \\
&= \psi_3(z) \sum_\alpha F_\alpha(w_i) \\
&\approx \psi_3(z) \sum_\alpha e^{-\frac{c}{6}f_\alpha(w_i)}\end{aligned}$$
(37)

Inserting above equation into (36), we get:

$$\left(\sum_{i=1}^4 \left(\frac{2h_i}{(z-w_i)^3} - \frac{1}{(z-w_i)^2}\partial_{w_i}\right) + \xi \partial_z \sum_{i=1}^4 \left(\frac{h_i}{(z-w_i)^2} - \frac{1}{(z-w_i)}\partial_{w_i}\right) + \eta \partial_z^3\right)\psi_3(z)\sum_\alpha e^{-\frac{c}{6}f_\alpha(w_i)} = 0$$
(38)

As we mentioned in the standard monodromy method, each conformal block $F_\alpha(w_i)$ needs to have its own monodromy condition that depends on the conformal dimension of itself h$\alpha$. As a result, each conformal block satisfies the differential Eq. (38) separately. So we get:

$$\left(\sum_{i=1}^4 \left(\frac{2h_i}{(z-w_i)^3} - \frac{1}{(z-w_i)^2}\partial_{w_i}\right) + \xi \partial_z \sum_{i=1}^4 \left(\frac{h_i}{(z-w_i)^2} - \frac{1}{(z-w_i)}\partial_{w_i}\right) + \eta \partial_z^3\right)\psi_3(z)e^{-\frac{c}{6}f_\alpha(w_i)} = 0$$
(39)

Similar to $\hat{\psi}_2$, $\hat{\psi}_3$ scales like $O(c^0)$, so we can neglect $\partial_{w_i}$ derivatives acting on $\hat{\psi}_3(z)$. We simplify the equation and finally get:

$$\psi_3'''(z) + \frac{c\xi}{6\eta}\sum_{i=1}^4\left(\frac{\varepsilon_i}{(z-w_i)^2} - \frac{c_i}{(z-w_i)}\right)\psi_3'(z) + \frac{c(\xi+1)}{6\eta}\sum_{i=1}^4\left(\frac{2\varepsilon_i}{-(z-w_i)^3} + \frac{c_i}{(z-w_i)^2}\right)\psi_3(z) = 0$$
(40)

Under the large $c$ limit we can get rid of $\zeta$ and $\eta$ and have:

$$\psi_3'''(z) + 4T(z)\psi_3'(z) + 2T'(z)\psi_3(z) = 0$$
(41)

where:

$$T(z) = \sum_{i=1}^{4}\left(\frac{\varepsilon_i}{(z-w_i)^2} - \frac{c_i}{(z-w_i)}\right), T'(z) = \sum_{i=1}^{4}\left(\frac{2\varepsilon_i}{-(z-w_i)^3} + \frac{c_i}{(z-w_i)^2}\right) \quad (42)$$

So we get the differential equation of the level-three monodromy method (41), which is a third order Fuchsian differential equation, see (104) in the Appendix. This is the differential equation based on which the level-three variant of the monodromy method is formulated. The stress tensor $T(z)$ and it's derivative which are meromorphic functions with four regular singularities at the points of insertion operators $w_i$, for $i = 1, ..., 4$ on the complex plane. Similar to the standard case, we can use the global conformal transformation and the regularity condition (20) to push all kinematic information of $w_i$ into a single conformal ratio $x = \frac{(w_1-w_2)(w_3-w_4)}{(w_1-w_3)(w_2-w_4)}$, then we obtain the final expressions:

$$T(z) = \frac{\varepsilon_1}{z^2} + \frac{\varepsilon_2}{(z-x)^2} + \frac{\varepsilon_3}{(z-1)^2} + \frac{\varepsilon_1+\varepsilon_2+\varepsilon_3-\varepsilon_4}{z(1-z)} - \frac{c_2(x)(1-x)x}{z(z-x)(1-z)}$$

$$T'(z) = -\frac{2\varepsilon_1}{z^3} - \frac{2\varepsilon_2}{(z-x)^3} - \frac{2\varepsilon_3}{(z-1)^3} + \frac{(2z|-1)(\varepsilon_1+\varepsilon_2+\varepsilon_3-\varepsilon_4)}{(z(z-1))^2} - \frac{c_2(x)(1-x)x(3z^2-2(1+x)z+x)}{(z(z-x)(1-z))^2}$$

(43)

**Monodromy condition.** Now we derive the monodromy condition for computing conformal blocks in this setup. This monodromy condition can be derived from the differential Eq. (36) for $\langle O_1 O_2 \hat{\psi}(z) O_3 O_4 \rangle$ (Similar to $\hat{\psi}_2(z)$, we omitted the subscript "3" in $\hat{\psi}_3(z)$ for conciseness). As we did in the standard monodromy method, by (23): $\psi(z) = \frac{\langle 0|O_\alpha(y)\hat{\psi}(z)O_3O_4|0\rangle}{\langle \alpha|O_3O_4\rangle}$, we can take the leading order of the OPE: $O_\alpha(y)\hat{\psi}(z) \sim (z-y)^k O_d(y)$ as $z \to y$, so we have:

$$\psi(z) = (z-y)^k \frac{\langle 0|O_d(y)O_3O_4|0\rangle}{\langle \alpha|O_3O_4\rangle} \sim (z-y)^k \quad (44)$$

Inserting $\psi(z)$ into the differential Eq. (36), we have:

$$\left(\eta\partial_z^3 + \xi\partial_z\left(\frac{h_\alpha}{(z-y)^2} + \frac{1}{z-y}\partial_y\right) - \frac{2h_\alpha}{(z-y)^3} - \frac{1}{(z-y)^2}\partial_y + \xi\partial_z \sum_{i=3,4}\left(\frac{h_i}{(z-w_i)^2}\right.\right.$$
$$\left.\left.+\frac{1}{z-w_i}\partial_{w_i} - \frac{2h_i}{(z-w_i)^3} - \frac{1}{(z-w_i)^2}\partial_{w_i}\right)\right)(z-y)^k \langle O_d(y)O_3O_4\rangle = 0$$

(45)

Then we use the limit $z \to y$, so we have $|z-w_i| \gg |z-y|$. It is straightforward to get:

$$\left(\eta\partial_z^3 + \xi\partial_z\left(\frac{h_\alpha}{(z-y)^2} + \frac{1}{z-y}\partial_y\right) - \frac{2h_\alpha}{(z-y)^3} - \frac{1}{(z-y)^2}\partial_y\right)(z-y)^k = 0 \quad (46)$$

This amounts to an equation for $k$, and we can find three solutions:

$$\begin{cases} k_1 = -h_{\psi_3} \\ k_2 = \frac{1}{2}\left(1 - h_{\psi_3} - \sqrt{1 - 2h_{\psi_3} + h_{\psi_3}^2 + 8h_\alpha + 8h_{\psi_3}h_\alpha}\right) \\ k_3 = \frac{1}{2}\left(1 - h_{\psi_3} + \sqrt{1 - 2h_{\psi_3} + h_{\psi_3}^2 + 8h_\alpha + 8h_{\psi_3}h_\alpha}\right) \end{cases} \quad (47)$$

And so we obtain the solutions:

$$\psi_{1,2,3}(z) \sim (z|-y)^{k_{1,2,3}} \quad (48)$$

where the subscripts 1, 2 and 3 stand for three different solutions. Again these exponents $k_i, i = 1, 2$ reflect the fusion rule between the level-three degenerate operator $\hat{\psi}$ and $O_\alpha$: $\hat{\psi} \times O_\alpha = O_{\beta_1} + O_{\beta_2} + O_{\beta_3}$, where the conformal dimesions of $O_{\beta_1}$, $O_{\beta_2}$ and $O_{\beta_3}$ are $h_\alpha - \frac{\sqrt{1-c}+\sqrt{25-c}}{\sqrt{6}}$, $h_\alpha$ and $h_\alpha + \frac{\sqrt{1-c}+\sqrt{25-c}}{\sqrt{6}}$, respectively. Similar to the standard case, we take $z$ to go through a contour that encircles $y$, i.e. $(z-y) \to (z-y)e^{2\pi i}$. In the OPE limit $\{w_1, w_2\} \to y$, this amounts to a contour that also encloses both $w_1$ and $w_2$. Then we have:

$$\begin{pmatrix} \psi_1 \\ \psi_2 \\ \psi_3 \end{pmatrix} \to M_3 \begin{pmatrix} \psi_1 \\ \psi_2 \\ \psi_3 \end{pmatrix} \quad (49)$$

where $M_3$ is the monodromy matrix of the level-three monodromy method:

$$M_3 = \begin{pmatrix} e^{2\pi i k_1} & 0 & 0 \\ 0 & e^{2\pi i k_2} & 0 \\ 0 & 0 & e^{2\pi i k_3} \end{pmatrix} \quad (50)$$

In the basis that diagonalizes the matrix. So the trace of the monodromy matrix is:

$$TrM_3 = e^{2\pi i k_1} + e^{2\pi i k_2} + e^{2\pi i k_3} = e^{-i\pi h_{\psi_3}}\left(e^{-i\pi h_{\psi_3}} - 2\cos\left(\pi\sqrt{8(h_{\psi_3}+1)h_\alpha + (h_{\psi_3}-1)^2}\right)\right) \quad (51)$$

Analogous to the standard case, the monodromy condition (51) combined with the differential Eq. (41) are used to in principle fix the accessory parameter $c_2(x)$, which can then be integrated to obtain the conformal block. The whole procedure constitutes the formulation of the monodromy method associated with the level-three degenerate operator. In particular, for the vacuum block, the monodromy matrix is the trivial identity:

$$M_3 = \begin{pmatrix} 1 & 0 & 0 \\ 0 & 1 & 0 \\ 0 & 0 & 1 \end{pmatrix} \quad (52)$$

Whose trace is 3. In the next section, we focus on computing the vacuum blocks using the level-three variant of the monodromy method, and benchmark it by checking against the standard level-two method.

## Computing vacuum block using level-three monodromy method

Having derived the level-three variant of the monodromy method, in this section we shall apply it to calculate conformal blocks. For simplicity we shall focus on computing the vacuum blocks, but in principle it can be applied to other blocks as well. Technically this involves solving $c_2(x)$ from the monodromy condition. As mentioned before in the case of the standard monodromy method, we should think of the monodromy matrix $M_3$ and therefore its trace $TrM_3$ both as functions of the accessory parameter $c_2(x)$ and the remaining parameters $\{\varepsilon_i, x\}$. Unfortunately, explicit expressions for these functions have not been found, and it is very plausible that the true answer involves highly transcendental functions of $c_2(x)$ as well as the remaining parameters. Solving the monodromy condition analytically is therefore beyond the current scope of monodromy methods, both for the standard case and the variant we have derived in "Level-three monodromy method" section.

We should therefore aim to solve the monodromy problem either approximately or numerically. The main purpose is to compare it against the corresponding results from the standard level-two monodromy method. By doing this we can both check the validity of the formulation, and also demonstrate explicitly the connection between solving the level-three and level-two monodromy problems. We emphasize again that although the connection may look obvious from a physics perspective: they aim to compute the same conformal block, at the technical level the two monodromy problems are not connected in any obvious manner, e.g. one cannot simply map from one into another. It is therefore a non-trivial check from a mathematical point of view regarding their mutual consistency. We shall comment more on this point in the "Discussion" section. The plan in this section is to first solve the level-three monodromy problem using perturbation theory in the heavy-light limit, and then to perform numerical calculations for generic cases. Both approaches will be compared against the corresponding results obtained using the standard level-two monodromy method.

**Perturbative calculation in the heavy-light limit.** Let us now apply the level-three monodromy method to compute the vacuum conformal block in the heavy-light limit perturbatively. The corresponding results were obtained using the standard monodromy method in[21,29], which we compare our result against to. As mentioned before, in the heavy-light limit we are interested in the case where the conformal dimensions $\varepsilon_i$ satisfy:

$$\varepsilon_L = \varepsilon_1 = \varepsilon_2 \ll 1, \quad \varepsilon_H = \varepsilon_3 = \varepsilon_4 \gg 1 \quad (53)$$

This is called the Heavy-Light (HL) limit because two of the external operators $O_1$ and $O_2$ have small conformal dimensions (so they are "light operators") while two of the remaining external operators $O_3$ and $O_4$ have much larger conformal dimension (so they are "heavy operators"). Through state-operator correspondence these correlation functions become expectation values of probe operators evaluated in high energy eigenstates, and therefore can be used to study the phenomena of Eigenstate Thermalization Hypothesis (ETH) in 2d CFTs[36–38]. To proceed, we treat the light operator's conformal dimension $\varepsilon_L$ as a small parameter and expand all the relevant parameters in the differential Eq. (41) in series expansions of $\varepsilon_L$:

$$\begin{aligned}
\psi(z) &= \psi^{(0)}(z) + \varepsilon_L \psi^{(1)}(z) + \varepsilon_L^2 \psi^{(2)}(z) + \cdots \\
c_2(x) &= c_2^{(0)}(x) + \varepsilon_L c_2^{(1)}(x) + \varepsilon_L^2 c_2^{(2)}(x) + \cdots \\
T(z) &= T^{(0)}(z) + \varepsilon_L T^{(1)}(z) + \varepsilon_L^2 T^{(2)}(z) + \cdots \\
T^{(0)}(z) &= \frac{\varepsilon_H}{(1-z)^2} \\
T^{(1)}(z) &= \frac{1}{(z-x)^2} + \frac{1}{z^2} + \frac{2}{z(1-z)} - \frac{c_2^{(0)}(x)(1-x)x}{\varepsilon_L z(z-x)(1-z)} \\
T^{(n)}(z) &= -\frac{c_2^{(n-1)}(x)x(1-x)}{\varepsilon_L z(z-x)(1-z)}, \quad n \geq 2
\end{aligned} \quad (54)$$

In this way, we decompose the monodromy problem in orders of $\varepsilon_L$, and the solution involving both $\psi(z)$ and $c_2(x)$ can be obtained order by order iteratively, at least in principle. Now we begin with the leading order differential equation for $\psi^{(0)}(z)$:

$$\psi'''(0)(z) + 4T^{(0)}(z)\psi'^{(0)}(z) + 2T'(0)(z)\psi^{(0)}(z) = 0 \tag{55}$$

It admits three independent solutions:

$$\begin{cases} \psi_1^{(0)}(z) = 1 - z \\ \psi_2^{(0)}(z) = (1-z)^{1+\sqrt{1-4\varepsilon_H}} \\ \psi_3^{(0)}(z) = (1-z)^{1-\sqrt{1-4\varepsilon_H}} \end{cases} \tag{56}$$

To simplify notations, it is useful to rewrite the monodromy Eq. (41) as a first-order matrix Ordinary Differential Equation (ODE):

$$\partial_z \Psi = a(z) \Psi \tag{57}$$

where

$$\Psi = \begin{pmatrix} (\partial_z^2 + 2T(z))\psi_1 & (\partial_z^2 + 2T(z))\psi_2 & (\partial_z^2 + 2T(z))\psi_3 \\ \partial_z \psi_1 & \partial_z \psi_2 & \partial_z \psi_3 \\ \psi_1 & \psi_2 & \psi_3 \end{pmatrix} \tag{58}$$

$$a(z) = \begin{pmatrix} 0 & -2T(z) & 0 \\ 1 & 0 & -2T(z) \\ 0 & 1 & 0 \end{pmatrix} \tag{59}$$

where in subscripts 1, 2, and 3 stand for three different solutions. Or another concise expression:

$$\Psi = \begin{pmatrix} \psi_1 & \psi_2 & \psi_3 \\ \partial_z \psi_1 & \partial_z \psi_2 & \partial_z \psi_3 \\ \partial_z^2 \psi_1 & \partial_z^2 \psi_2 & \partial_z^2 \psi_3 \end{pmatrix} \tag{60}$$

$$a(z) = \begin{pmatrix} 0 & 1 & 0 \\ 0 & 0 & 1 \\ -2T'(z) & 4T(z) & 0 \end{pmatrix} \tag{61}$$

We want to solve the ODE perturbatively, so we can split $\Psi$, $a(z)$ as:

$$\Psi = \Psi^{(0)} \Psi^{(1)}$$
$$a(z) = a^{(0)}(z) + a^{(1)}(z) \tag{62}$$

where

$$\Psi^{(0)} = \begin{pmatrix} (\partial_z^2 + 2T^{(0)}(z))\psi_1^{(0)} & (\partial_z^2 + 2T^{(0)}(z))\psi_2^{(0)} & (\partial_z^2 + 2T^{(0)}(z))\psi_3^{(0)} \\ \partial_z \psi_1^{(0)} & \partial_z \psi_2^{(0)} & \partial_z \psi_3^{(0)} \\ \psi_1^{(0)} & \psi_2^{(0)} & \psi_3^{(0)} \end{pmatrix} \tag{63}$$

$$a^{(0)}(z) = \begin{pmatrix} 0 & -2T^{(0)}(z) & 0 \\ 1 & 0 & -2T^{(0)}(z) \\ 0 & 1 & 0 \end{pmatrix} \tag{64}$$

$$a^{(1)}(z) = a(z) - a^{(0)}(z) = \begin{pmatrix} 0 & -2\varepsilon_L T^{(1)}(z) & 0 \\ 0 & 0 & -2\varepsilon_L T^{(1)}(z) \\ 0 & 0 & 0 \end{pmatrix} + O(\varepsilon_L^2) \tag{65}$$

And hence:

$$\partial_z \Psi^{(0)} = a^{(0)}(z) \Psi^{(0)} \tag{66}$$

With the help of (57), (62) and (66), we can know the differential equation on $\Psi^{(1)}$ is:

$$\partial_z \Psi^{(1)} = \left(\Psi^{(0)}\right)^{-1} a^{(1)}(z) \Psi^{(0)} \Psi^{(1)} \tag{67}$$

So the solution at linear order in $\varepsilon_L$ is:

$$\Psi^{(1)} = P \exp\left(\int dz \left(\Psi^{(0)}\right)^{-1} a^{(1)}(z) \Psi^{(0)}\right)$$
$$= 1 + \int dz \left(\Psi^{(0)}\right)^{-1} a^{(1)}(z) \Psi^{(0)} + O(\varepsilon_L^2) \tag{68}$$

Our goal is to compute the vacuum block, which means we need to impose the trivial monodromy condition $M^{(1)}_{\gamma=\{0,x\}} = 0$, where the $\gamma = \{0, x\}$ represents a contour enclosing the poles at $z = 0$ and $z = x$ in the complex plane. To the leading order in $\varepsilon_L$, this requires that:

$$\oint_{\gamma=\{0,x\}} dz \left(\Psi^{(0)}\right)^{-1} a^{(1)}(z) \Psi^{(0)} = 0 \tag{69}$$

Using the expression of $\Psi^{(0)}$ and $a^{(1)}(z)$, we get the expression of the integrand:

$$\left(\Psi^{(0)}\right)^{-1} a^{(1)}(z) \Psi^{(0)} = \begin{pmatrix} 0 & -\frac{2T^{(1)}(z)(1-z)^{\alpha_H+1}}{\alpha_H} & \frac{2T^{(1)}(z)(1-z)^{1-\alpha_H}}{\alpha_H} \\ \frac{T^{(1)}(z)(1-z)^{1-\alpha_H}}{\alpha_H} & -\frac{2T^{(1)}(z)(z-1)}{\alpha_H} & 0 \\ \frac{T^{(1)}(z)(z-1)(1-z)^{\alpha_H}}{\alpha_H} & 0 & \frac{2T^{(1)}(z)(z-1)}{\alpha_H} \end{pmatrix} \tag{70}$$

where $\alpha_H \equiv \sqrt{1 - 4\varepsilon_H}$. The condition (69) requires that the sum of residues of around 0 and $x$ vanish. We calculate these residues and find:

$$\begin{aligned}
&Res_{z=0}\left(-\frac{2T^{(1)}(z)(1-z)^{\alpha_H+1}}{\alpha_H}\right) + Res_{z=x}\left(-\frac{2T^{(1)}(z)(1-z)^{\alpha_H+1}}{\alpha_H}\right) \\
&= \frac{2\left((1-x)^{\alpha_H}\left(-c_2^{(0)}x + c_2^{(0)} + \alpha_H \varepsilon_L + \varepsilon_L\right) + c_2^{(0)}x - c_2^{(0)} + \alpha_H \varepsilon_L - \varepsilon_L\right)}{\alpha_H} \\
&Res_{z=0}\left(\frac{2T^{(1)}(z)(1-z)^{1-\alpha_H}}{\alpha_H}\right) + Res_{z=x}\left(\frac{2T^{(1)}(z)(1-z)^{1-\alpha_H}}{\alpha_H}\right) \\
&= \frac{2\left((1-x)^{-\alpha_H}\left(c_2^{(0)}x - c_2^{(0)} + \alpha_H \varepsilon_L - \varepsilon_L\right) - c_2^{(0)}x + c_2^{(0)} + \alpha_H \varepsilon_L + \varepsilon_L\right)}{\alpha_H} \\
&Res_{z=0}\left(\frac{2T^{(1)}(z)(1-z)}{\alpha_H}\right) + Res_{z=x}\left(\frac{2T^{(1)}(z)(1-z)}{\alpha_H}\right) = 0 \\
&Res_{z=0}\left(\frac{T^{(1)}(z)(z-1)(1-z)^{\alpha_H}}{\alpha_H}\right) + Res_{z=x}\left(\frac{T^{(1)}(z)(z-1)(1-z)^{\alpha_H}}{\alpha_H}\right) \\
&= \frac{(1-x)^{-\alpha_H}\left(c_2^{(0)}x - c_2^{(0)} + \alpha_H \varepsilon_L - \varepsilon_L\right) - c_2^{(0)}x + c_2^{(0)} + \alpha_H \varepsilon_L + \varepsilon_L}{\alpha_H}
\end{aligned} \tag{71}$$

Demanding that they vanish yields:

$$c_2^{(0)}(x) = \frac{(1+\alpha_H)(1-x)^{\alpha_H} - 1 + \alpha_H}{(1-x)((1-x)^{\alpha_H} - 1)} \varepsilon_L \tag{72}$$

This result is the same as that calculated in[21,29], which uses the standard monodromy method. We conclude from this analysis that to leading order in perturbation theory, the level-three monodromy method is analytically consistent with the standard level-two monodromy method. In principle, we can check subleading corrections, but in practice this becomes immediately intractable beyond the leading order, for example, see[39]. We should therefore refrain from going to higher orders in $\varepsilon_L$ and resort to numerical calculations for checks beyond perturbative values for $\varepsilon_L$.

**Numerical solutions for generic values of $\varepsilon_i$.** In this section, we present some numerical results for solving the level-three monodromy problem. For those who might be interested we first briefly explain the numerical procedure for solving the monodromy problem. The crucial step is to compute for a given set of parameters $\{\varepsilon_i, x, c_2\}$ the monodromy matrix $M_3$. In practice we decompose it into four parts $M_3 = M_0 R_{0,x} M_x R_{0,x}^{-1}$ so that each part can be computed relatively easily, see Fig. 3.

In particular, the diagonal matrices $M_0$ and $M_x$ encode the local phases that the three power-law solutions $\{z^{s1,2,3}\}$ and $\{(z-x)^{s1,2,3}\}$ obtain around $z = 0$ and $z = x$ respectively, where $s_1 = 1, s_2 = 1 + \sqrt{1-4\varepsilon_H}, s_3 = 1 - \sqrt{1-4\varepsilon_H}$:

$$M_0 = M_x = \begin{pmatrix} e^{2\pi i} & 0 & 0 \\ 0 & e^{2\pi i(1+\sqrt{1-4\varepsilon_H})} & 0 \\ 0 & 0 & e^{2\pi i(1-\sqrt{1-4\varepsilon_H})} \end{pmatrix} \tag{73}$$

While the "Scattering Matrix" $R_{0,x}$ denotes how the modes $\{z^{s1,2,3}\}$ scatter into the modes $\{(z-x)^{s1,2,3}\}$ as they "Propagate" via the differential Eq. (41) from near $z = 0$ to near $z = x$:

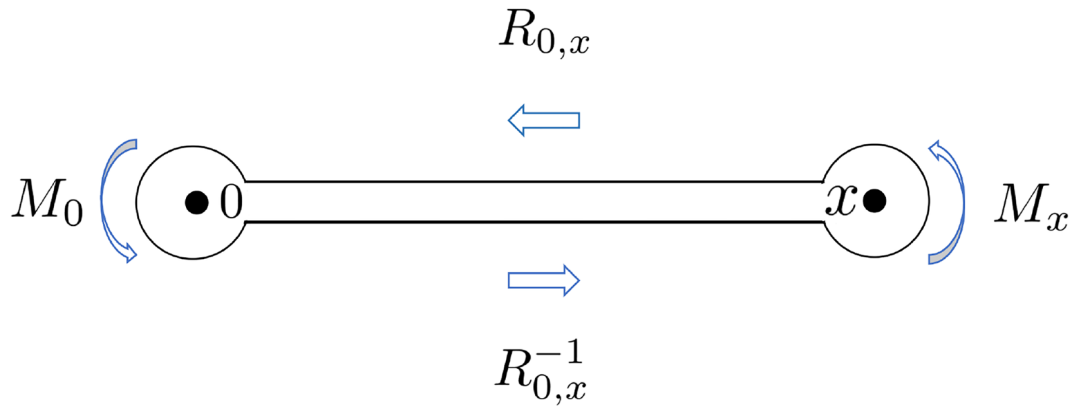

**Figure 3.** We decompose the monodromy matrix into four parts $M_3 = M_0 R_{0,x} M_x R_{0,x}^{-1}$, where matrices $M_0$ and $M_x$ encode the local phases that three solutions $\{z^{s1,2,3}\}$ and $\{(z-x)^{s1,2,3}\}$ obtain around $z=0$ and $z=x$ respectively; while the "Scattering Matrix" $R_{0,x}$ represents how the modes $\{z^{s1,2,3}\}$ scatter into the modes $\{(z-x)^{s1,2,3}\}$ as they "Propagate" via the differential Eq. (41) from near $z=0$ to near $z=x$. We need to make the radii very small for this representation to be accurate.

$$\psi_i(z \sim 0) \to \sum_{j=1}^{3} R_{0,x}^{i,j} \psi_j(z \sim x), \quad i = 1, 2, 3 \tag{74}$$

where

$$\psi_i(z \sim 0) \approx z^{s_i}, \quad \psi_i(z \sim x) \approx (z-x)^{s_i} \tag{75}$$

It requires numerically integrating the differential equation and then solving the matrix elements $R_{0,x}^{i,j}$ by imposing matching conditions for $\{\psi, \psi', \psi''\}$ near $z = x$. The accessory parameters $c_2(x)$ are then obtained by essentially applying Newton's interation method to the monodromy equation:

$$Tr M_3 = 3 \tag{76}$$

Starting close to $x_{n=0} \approx 0$, where we have a good guess for $c_2(x_0)$ using the OPE[40]:

$$c_2(x_0) \approx \frac{2\varepsilon_L}{x_0} - \frac{2|}{3} x_0 \varepsilon_H \varepsilon_L + x_0^3 \left( \frac{22\varepsilon_H^2 \varepsilon_L^2}{135} - \frac{2\varepsilon_H^2 \varepsilon_L}{45} - \frac{2\varepsilon_H \varepsilon_L^2}{45} - \frac{6\varepsilon_H \varepsilon_L}{5} \right) + x_0^4 \left( \frac{11\varepsilon_H^2 \varepsilon_L^2}{27} - \frac{\varepsilon_H^2 \varepsilon_L}{9} - \frac{\varepsilon_H \varepsilon_L^2}{9} - \frac{4\varepsilon_H \varepsilon_L}{3} \right) \tag{77}$$

And then interatively use $c_2(x_n)$ as a guess to find the solution for $c_2(x_{n+1})$, where $x_{n+1}$ is a nearby point of $x_n$ along some trajectory in complex $x$ plane that we want to compute the values of the vacuum block $F_0(x)$.

So by numerically solving the monodromy problem rather than perturbatively, we can compare the value of accessory parameter $c_2(x) = -\frac{6}{c} \partial_x \ln F_0(x)$ for non-perturbative values $\varepsilon_L = 1, 2, 3, 4$ obtained from the two monodromy methods. See Figs. 4 and 5 for details.

### Generalizing to higher-level monodromy method

We have derived and checked the explicit formulation of the monodromy problem associated with the level-three degenerate operator, we see that it is consistent with the standard level-two monodromy method. In this section, we make a further generalization and consider the consequence of inserting a more general degenerate operator $\hat{\psi}_r$ into the correlation function. In particular, we focus on those $\hat{\psi}_r$ that are associated with the Verma modules $V_{r,1}$, i.e. they contain null states at level $r$. A common expression of conformal dimension of $\hat{\psi}_r$ is[41]:

$$h_{r,1}(t) = \frac{1}{4}(r^2 - 1)t - \frac{1}{2}(r - 1) \tag{78}$$

where

$$t = 1 + \frac{1}{12}(1 - c + \sqrt{(c-1)(c-25)}) \tag{79}$$

In the semi-classical limit $c \to \infty$, the conformal dimension of $\psi_r$ is:

$$\lim_{c \to \infty} h_{r,1} = \frac{1-r}{2} \tag{80}$$

The null state is obtained by acting a covariant differential operator $\Delta_{r,1}$ of degree $r$ on the state $|\psi_r\rangle$, which is given by:

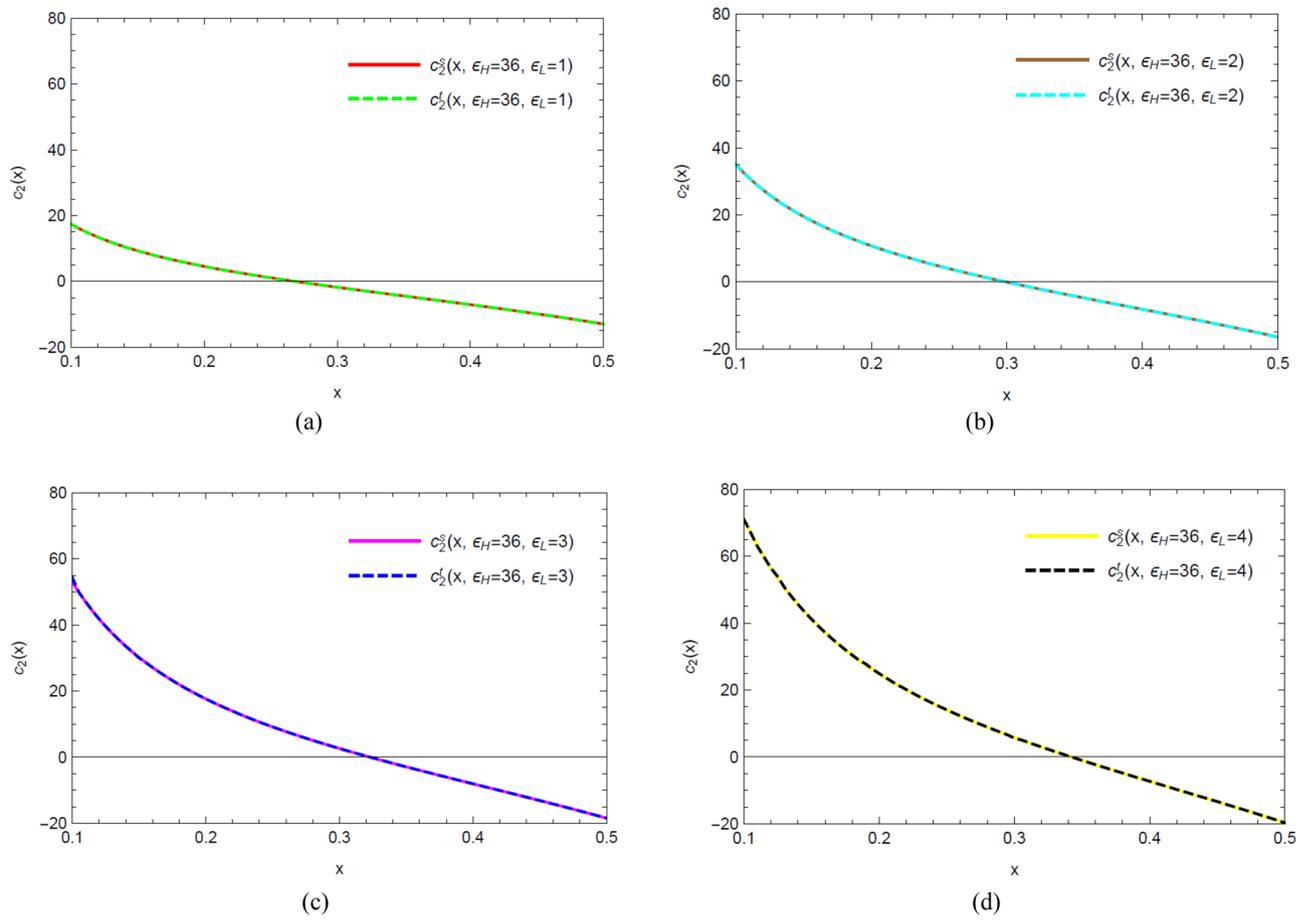

**Figure 4.** Plots of accessory parameters $c_2(x)$ of the large $c$ vacuum Virasoro blocks from the standard monodromy method and the level-three monodromy method, as conformal dimensions $\varepsilon_H = 36$ and $\varepsilon_L$ varies from 1 to 4. Pay attention that the values of $\varepsilon_L$, which are beyond the perturbative values $\varepsilon_L \ll 1$, scale like $O(1)$. $c_2^s(x)$ denotes the accessory parameter obtained from the standard method while $c_2^t(x)$ denotes the one obtained from the level-three method. The overlap of the trajectories of $c_2^s(x)$ and $c_2^t(x)$ shows that these two methods are numerically in agreement with each other.

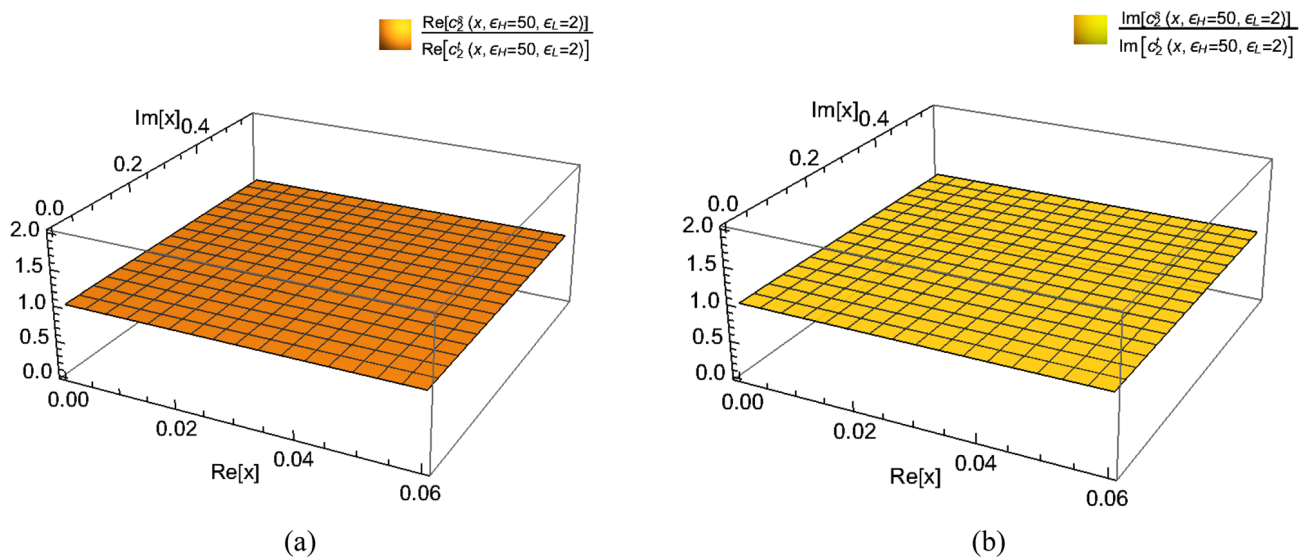

**Figure 5.** Plots of the ratio of real (imaginary) part of accessory parameter $c_2^s(x)$ to real (imaginary) part of accessory parameter $c_2^t(x)$ where $x$ is a complex variable, as conformal dimensions $\varepsilon_H = 50$ and $\varepsilon_L = 2$. The horizontal plane very closing to 1 indicates the consistency between the two methods.

$$|\chi_r\rangle = \Delta_{r,1}(t)|\psi_r\rangle = \det\left[-J_- + \sum_{m=0}^{\infty}(-tJ_+)^m L_{-m-1}\right]|\psi_r\rangle$$
$$= \sum_{p_i \geq 1, p_1+\cdots+p_i=r} \frac{[(r-1)!]^2(-t)^{r-k}}{\prod_{i=1}^{k-1}(p_1+\cdots+p_i)(r-p_1-\cdots-p_i)} L_{-p_1}\ldots L_{-p_k}|\psi_r\rangle \quad (81)$$

where

$$[J_0]_{i,j} = \frac{1}{2}(r-2i+1)\delta_{i,j}$$
$$[J_-]_{i,j} = \begin{cases} \delta_{i,j+1}, & j=1,2,\ldots,r-1 \\ 0, & j=r \end{cases}$$
$$[J_+]_{i,j} = \begin{cases} i(r-i)\delta_{i+1,j}, & i=1,2,\ldots,r-1 \\ 0, & i=r \end{cases} \quad (82)$$

So by inserting $\hat{\psi}_r$ into the four-point function, we have a differential equation of order $r$:

$$\gamma_{r,1}(z,\partial_z,w_i,\partial_{w_i})\langle O_1(w_1)O_2(w_2)\hat{\psi}_r(z)O_3(w_3)O_4(w_4)\rangle = 0 \quad (83)$$

where

$$\gamma_{r,1}(z,\partial_z,w_i,\partial_{w_i}) = \det\left[-J_- + \sum_{m=0}^{\infty}(-tJ_+)^m L_{-m-1}\right]$$
$$= \sum_{p_i \geq 1, p_1+\cdots+p_i=r} \frac{[(r-1)!]^2(-t)^{r-k}}{\prod_{i=1}^{k-1}(p_1+\cdots+p_i)(r-p_1-\cdots-p_i)} L_{-p_1}\ldots L_{-p_k} \quad (84)$$

With:

$$L_{-1} = \partial_z, \quad L_{-m} = \sum_i \left(\frac{(m-1)h_i}{(w_i-z)^m} - \frac{1}{(w_i-z)^{m-1}}\partial_{w_i}\right), \quad m \geq 2 \quad (85)$$

Assuming again that the degenerate operator is light, i.e. $r \ll c$, one can still invoke the semi-classical approximation of the five-point functions:

$$\langle O_1(w_1)O_2(w_2)\hat{\psi}_r(z)O_3(w_3)O_4(w_4)\rangle \approx \psi_r(z)\sum_\alpha e^{-\frac{c}{6}f_\alpha(w_i)} \quad (86)$$

By the same logic as before, each block needs to satisfy the differential equation separately. We thus obtain that:

$$\gamma_{r,1}(z,\partial_z,w_i,\partial_{w_i})\left(\psi_r(z)e^{-\frac{c}{6}f_\alpha(w_i)}\right) = 0 \quad (87)$$

In this paper, we are only interested in computing four-point conformal blocks, for which we can use the global conformal transformation to reduce the dependence on kinematic data $\{w_i\}$ to a single conformal ratio $x$, as we did before. We mention for the purpose of generality that one can also consider the monodromy method associated with higher-point conformal blocks, which can be derived by inserting a degenerate operator analogous to the four-point conformal blocks. When there are $n \geq 4$ operators in the correlator, each additional operator corresponds to an additional conformal-invariant, as a result the corresponding monodromy method involves $n$-3 accessory parameters solving $n$-3 monodromy conditions, corresponding to $n$-3 OPE processes that specify a channel[32]. For instance, the monodromy method for conformal blocks of two heavy operators and an arbitrary number of light operators has been illustrated in the work[33]. The paper[42] discussed the computation of 5-point conformal blocks with two heavy, two light, and one superlight operator at the large central charge using the monodromy method. In the works[30,43], the authors showed that any point conformal blocks in the semiclassical limit can be computed by solving a monodromy problem similar to the standard monodromy method we discussed in "Standard monodromy method" section.

Similar to the level-two and level-three cases, we can pack the $r$-th order differential equation into a first-order matrix form. Upon observation, one can check that the general form of the differential equation takes the form:

$$\psi_r^{(n)}(z) + p_2(z)\psi_r^{(n-2)}(z) + \cdots + p_n(z)\psi_r(z) = 0 \quad (88)$$

where $p_{2,\ldots,n}(z)$ are the polynomials that have regular singular points in the complex domain. There is a curious feature that the $\psi_r^{(n-1)}(z)$ terms are always absent, it is a consequence of the special structure associated with the differential operator (84). In terms of a first-order matrix differential Eq. (88) can be written as:

$$\partial_z \Psi_r = A(z) \Psi_r \quad (89)$$

where $\Psi_r$ can be defined by:

$$\Psi_r = \begin{pmatrix} \psi_r \\ \psi_r' \\ \psi_r'' \\ \vdots \\ \psi_r^{(n-1)} \end{pmatrix} \quad (90)$$

In this basis, the matrix $A(z)$ is:

$$A(z) = \begin{pmatrix} 0 & 1 & 0 & \cdots & 0 & 0 \\ 0 & 0 & 1 & \cdots & 0 & 0 \\ \vdots & \vdots & & \ddots & \vdots & \vdots \\ 0 & 0 & \cdots & \cdots & 1 & 0 \\ 0 & 0 & \cdots & \cdots & 0 & 1 \\ -p_n(z) & -p_{n-1}(z) & \cdots & \cdots & -p_2(z) & 0 \end{pmatrix} \quad (91)$$

We can also derive the general monodromy condition via the fusion rule[41]:

$$\hat{\psi}_r \times O_\alpha = \sum_{k=1-r, k+r=1 \text{ mod } 2}^{k=r+1} O_{\alpha+k\alpha_+} \quad (92)$$

where the conformal dimensions of the operators are:

$$h_{r,1} = h_0 + \frac{1}{4}(r\alpha_+ + \alpha_-)^2, \quad h_\alpha = h_0 + \frac{1}{4}\alpha^2, \quad h_{\alpha+k\alpha_+} = h_0 + \frac{1}{4}(\alpha + k\alpha_+)^2 \quad (93)$$

With:

$$h_0 = \frac{1}{24}(c-1), \quad \alpha_\pm = \frac{\sqrt{1-c} \pm \sqrt{25-c}}{\sqrt{24}} \quad (94)$$

Recall that we did in deriving the level-two and level-three monodromy condition, we consider the three-point correlation function $\left\langle \hat{\psi}_r(z) O_\alpha(y) O_\beta\left(\frac{w_3+w_4}{2}\right) \right\rangle \sim (z-y)^{-(h_{r,1}+h_\alpha-h_\beta)}$ as $z$ encircles $y$, i.e. $\{w_1, w_2\}$, we can parametrize the conformal dimension of $O_\beta$ as:

$$h_\beta = h_0 + \frac{1}{4}\beta^2 \quad (95)$$

So by the fusion rule (92), we have:

$$\beta = \sum_{k=1-r, k+r=1 \text{ mod } 2}^{k=r-1} \alpha + k\alpha_+ \quad (96)$$

Reverting to the conformal dimensions notation, we have the exponent of the expression of the three-point correlation function:

$$-(h_{r,1} + h_\alpha - h_\beta) = \frac{1}{4}(k\alpha_+)^2 + \sqrt{h_\alpha - h_0}k\alpha_+ + h_{r,1} \quad (97)$$

where $k \in [1-r, r-1], \quad k+r = 1 \text{ mod } 2$

For the three-point correlation function, we take $z$ to go through a contour that encloses $y$: $(z-y) \to (z-y)e^{2\pi i}$ in the OPE limit $\{w_1, w_2\} \to y$. Then we have the general monodromy:

$$M_r = \begin{pmatrix} e^{i\pi \Lambda_{k_1}} & 0 & \cdots & 0 \\ 0 & e^{i\pi \Lambda_{k_2}} & \cdots & 0 \\ \vdots & \vdots & \ddots & \vdots \\ 0 & 0 & \cdots & e^{i\pi \Lambda_{k_r}} \end{pmatrix} \quad (98)$$

where $\Lambda_{k_i}$ are:

$$\Lambda_{k_1} = \frac{1}{2}((1-r)\alpha_+)^2 + 2\sqrt{h_\alpha - h_0}(1-r)\alpha_+ + 2h_{r,1}$$
$$\Lambda_{k_2} = \frac{1}{2}((3-r)\alpha_+)^2 + 2\sqrt{h_\alpha - h_0}(3-r)\alpha_+ + 2h_{r,1}$$
$$\dots$$
$$\Lambda_{k_r} = \frac{1}{2}((r-1)\alpha_+)^2 + 2\sqrt{h_\alpha - h_0}(r-1)\alpha_+ + 2h_{r,1}$$

(99)

Finally, we can solve the differential equation for the higher-level monodromy method (89) perturbatively and compute the conformal blocks (accessory parameters) by imposing corresponding monodromy conditions (98), just like the processes we took in "Computing vacuum block using level-three monodromy method" section.

We give the expression of the null state at level four and the conformal dimension of the degenerate operator:

$$|\chi_{4,1}\rangle = \left[L_{-1}^4 - 3tL_{-1}^2 L_{-2} + 12 L_{-1} L_{-3} + 9t^2 L_{-2}^2 - 36 t^3 L_{-4}\right]$$
$$|\psi_4\rangle h_{\psi_4} = \frac{15}{4}t - \frac{3}{2}$$

(100)

And we can also derive the differential equation of the level-four monodromy method and corresponding monodromy condition:

$$\psi_4^{(4)}(z) - \frac{c}{2} t T(z) \psi_4''(z) + 2c T'(z) \psi_4'(z) - 3c t^3 T''(z) \psi_4(z) + \sum_{i=1}^{4} \left( t^2 \frac{c(w_i - z)\left(c_i(c(w_i - z)c_i + 12 h_i + 6) + 6(z - w_i) c_i'(w_i)\right) + 36 h_i(h_i + 2)}{4(w_i - z)^4} \right) \psi_4(z) = 0$$

$$M_4 = \begin{pmatrix} e^{i\pi\left(\frac{9}{2}\alpha_+^2 - 6\sqrt{h_\alpha - h_0}\alpha_+ + 2h_{\psi_4}\right)} & 0 & 0 & 0 \\ 0 & e^{i\pi\left(\frac{1}{2}\alpha_+^2 - 2\sqrt{h_\alpha - h_0}\alpha_+ + 2h_{\psi_4}\right)} & 0 & 0 \\ 0 & 0 & e^{i\pi\left(\frac{1}{2}\alpha_+^2 + 2\sqrt{h_\alpha - h_0}\alpha_+ + 2h_{\psi_4}\right)} & 0 \\ 0 & 0 & 0 & e^{i\pi\left(\frac{9}{2}\alpha_+^2 + 6\sqrt{h_\alpha - h_0}\alpha_+ + 2h_{\psi_4}\right)} \end{pmatrix}$$

$$\mathrm{Tr} M_4 = e^{i\pi\left(2h_{\psi_4} + \frac{1}{2}\alpha_+^2\right)} \left(2\left(\cos\left(2\pi\sqrt{h_\alpha - h_0}\alpha_+\right) + \cos\left(6\pi\sqrt{h_\alpha - h_0}\alpha_+\right)\right) + e^{i\pi 4\alpha_+^2}\right)$$

(101)

## Discussions and conclusions

In this paper, we first reviewed the derivation of the standard method of monodromy. It involves inserting a level-two degenerate operator $\hat{\psi}_2$ into the four-point correlation function. In the semi-classical limit, the five-point function is expected to factorize into the original fourpoint function multiplying a "Wave Function" factor $\psi_2(z)$ of the degenerate operator (2). In this limit, the degeneracy condition of $\hat{\psi}_2$ gives rise to a second order differential equation for the factor $\psi(z)$. The differential equation is of Fuchsian type and contains regular singularities at the positions of operator insertions. It depends on parameters such as the external operator conformal dimensions $\varepsilon_i$ and an accessory parameter $c_2(x)$. Furthermore, $\hat{\psi}_2$ has truncated fusion relations with general Virasoro family $\alpha$, therefore the solutions to the differential equations should satisfy a monodromy condition consistent with the fusion relation if the accessory parameter $c_2(x)$ is related to the conformal block by $c_2(x) = -\frac{6}{c}\partial_x \ln F_\alpha(x)$. This provides the machinery for computing the conformal block: one uses the monodromy condition as an equation to solve the accessory parameter. In some loose sense, this approach bares certain similarities to the bootstrap philosophy in computing CFT quantities[44,45].

The main motivation of this paper is to apply the observation that we could have inserted any other degenerate operator into the correlation function at the beginning of the derivation. The particular choice of inserting the level-two degenerate operator has the advantage of optimizing the representation of the method: it involves solving differential equations of the lowest degree; from a formal perspective there is nothing special about the level-two degenerate operator. Therefore we investigated the consequences of inserting other degenerate operators into the four-point function and applying the ansatz (2) of the resulting five-point function in the semiclassical limit. The consequences give rise to variants of the standard method of monodromy. We explicitly formulated the method for a level-three degenerate operator, it takes the form of solving a third order differential equation and imposing the monodromy condition on the 3 × 3 monodromy matrix. We applied the method to compute vacuum conformal blocks using both perturbation theory and numerical method, as analytic solutions to the possibly transcendental monodromy equation are not currently feasible. The results were found to agree with the standard monodromy method. We also discussed the cases where a more general level-$r$ degenerate operator of the type $h_{r,1}$ is inserted into the correlation function. In this case, one needs to solve a $r$-th order differential equation and impose corresponding monodromy condition on the $r \times r$ monodromy matrix.

The freedom for choosing which degenerate operator to proceed the derivation has a few implications. They constitute the main messages we hope to discuss and convey in this paper, apart from the technical results. Firstly, for each choice of the degenerate operator at level $r \ll c$, we obtain a variant of the monodromy method that allows us to compute conformal blocks. Secondly, these methods are mutually consistent: they compute the same conformal blocks. This may sound like a logical tautology, but let us unpack what it means in practice. It means the following. For any particular variant of the monodromy method defined by a r-th order differential equation, for illustration we copy below the corresponding equations for $r = 2, 3, 4$:

$$\psi_2''(z) + T(z)\psi_2(z) = 0$$
$$\psi_3'''(z) + 4T(z)\psi_3'(z) + 2T'(z)\psi_3(z) = 0$$
$$\psi_4^{(4)}(z) - \frac{c}{2}tT(z)\psi_4''(z) + 2cT'(z)\psi_4'(z) - 3ct^3T''(z)\psi_4(z) + \sum_{i=1}^{4}\left(t^2\right.$$
$$\left.\frac{c(w_i - z)\big(c_i(c(w_i - z)c_i + 12h_i + 6) + 6(z - w_i)c_i'(w_i)\big) + 36h_i(h_i + 2)}{4(w_i - z)^4}\right)\psi_4(z) = 0$$
(102)

If the accessory parameter $c_2(x)$ is adjusted such that the $r$ solutions satisfy the monodromy condition associated with the computation of a particular conformal block $\alpha$, imposed on the $r \times r$ monodromy matrix, which for illustration we copy the traces for $r = 2, 3, 4$:

$$TrM_2 = -2\cos\left(\pi\sqrt{1 - \frac{24h_\alpha}{c}}\right)$$
$$TrM_3 = e^{-i\pi h_{\psi_3}}\left(e^{-i\pi h_{\psi_3}} - 2\cos\left(\pi\sqrt{8(h_{\psi_3} + 1)h_\alpha + (h_{\psi_3} - 1)^2}\right)\right)$$
$$TrM_4 = e^{i\pi\left(2h_{\psi_4} + \frac{1}{2}\alpha_+^2\right)}\left(2\left(\cos\left(2\pi\sqrt{h_\alpha - h_0}\alpha_+\right) + \cos\left(6\pi\sqrt{h_\alpha - h_0}\alpha_+\right)\right) + e^{i\pi 4\alpha_+^2}\right)$$
(103)

Then plugging the same $c_2(x)$ into other variants of the monodromy method will automatically solve the corresponding monodromy condition for conformal block $\alpha$. When viewed from this perspective, the mutual consistency condition predicts a non-trivial connection among monodromies of different types of Fuchsian systems that do not seem to be related in obvious ways. In principle, one could study these Fuchsian systems and verify the relation between their monodromy structures. Due to the author's lack of expertise in the subject, we refrain from making further comments. We hope to inspire further studies or explanations of this phenomenon, especially from mathematicians.

Instead, let us trace the physical origin of the connection. Recall that without making any approximations, the degeneracy conditions for distinct degenerate operators $\hat{\psi}_r$ are independent: the differential equations are satisfied by five-point functions $\left\langle O_1 O_2 \hat{\psi}_r(z) O_3 O_4 \right\rangle$ which are independent objects for different $\hat{\psi}_r$'s. Therefore at the exact level, there is no connection between their consequences. What ties them together is the assumption that the five-point functions factorize as (2) in the semi-classical limit for $r \ll c$ such that the degenerate operator behaves like a probe. From this observation we are led to the insight that the connections we are seeing between the Fuchsian systems that define different monodromy methods, in the form of mutual consistencies described before, serve to reflect the physical principle underlying the method, i.e. factorization (2) in the semi-classical limit. We can interpret this in two-folds: on the one hand, the mutual consistency connections between the Fuchsian systems appear non-trivial because they stem from an assumption that is inspired by physical principle; on the other hand, the fact that these connections can in principle be verified mathematically also implies that the assumption has a certain mathematical foundation in its truthfulness, where for example[21] has provided some scaling arguments for it.

It is interesting to discuss our results with other works involving conformal blocks. In the paper[46], the author provided a simple algebraic iterative method to compute the conformal blocks to any order in x, without taking the large central charge limit which we have taken in our paper. After taking the large c limit, the author found that the results agree with the one obtained by our standard monodromy method and level-three method which assume the exponentiation ansatz (4) of conformal blocks in the large central charge. The author also explored the monodromy method for the torus topology, gave the results for the classical blocks, and compared them with the conformal blocks under the large central charge limit in[47]. It would be useful to generalize our higher-level monodromy problems to these torus conformal blocks and discuss the rich algebraic structures behind them. Recently, in the work[48], the level-three degenerate operators arise in the monodromy problem in CFT with $W_3$ algebra. The reason why the level-three degenerate operator must appear here is rooted in the $W_3$ algebra, i.e. the commutation relations, whose generators $W_{-1}$ and $W_{-2}$ cannot simply act like $L_{-1} = \partial_z$, $L_{-m} = \sum_i \left(\frac{(m-1)h_i}{(w_i - z)^m} - \frac{1}{(w_i - z)^{m-1}} \partial_{w_i}\right)$, $m \geq 2$ by acting on $\psi(z)$. In this sense, it is reasonable to take the higher-level monodromy methods based on the Virasoro algebra and the monodromy method for the $W_3$ algebra blocks to be different implementation variants of the same idea[30]. This can be seen more clearly from the differential equation of the $W_3$ algebra monodromy method in[48], which differs from the third-order differential Eq. (41) we got by only one term contributed by the generator $W_{-3}$.

We proposing some future directions. Firstly, as mentioned before it would be interesting to further study the mathematical connections among the Fuchsian systems dictated by mutual consistencies among the monodromy methods, i.e. to verify or explain them. Secondly, from a practical point of view it would be interesting to explore whether there are problems for which other variants of the monodromy method are better suited. Related to this, it is also worth thinking about whether combining multiple monodromy methods can provide additional mileage in computing conformal blocks, for example by providing independent constraints mimicking the idea of CFT bootstrap. Next, it is curious to study what would happen as we increase the level $r$ of the degenerate operator, until it becomes semi-classical itself and the factorizable ansatz (2) breaks down; how does this breakdown manifest at the level of the connections we discovered among the Fuchsian systems. Last but not least, it would be useful to generalize our calculations and analyses to other types of conformal blocks such as $W_N$ conformal blocks[29], Neveu-Schwarz superconformal blocks[49], superconformal torus blocks[50], etc. There are different differential equations and rich algebraic structures that are worth exploring.

### Data availability
The datasets used and/or analyzed during the current study available from the corresponding author on reasonable request.

### References
1. Witten, E. Anti-de Sitter space and holography. *Adv. Theor. Math. Phys.* **2**, 253–291 (1998).
2. Gubser, S. S., Klebanov, I. R. & Polyakov, A. M. Gauge theory correlators from non-critical string theory. *Phys. Lett. B* **428**(1–2), 105–114 (1998).
3. Polchinski, J. Introduction to gauge/gravity duality. In *Theoretical Advanced Study Institute in Elementary Particle Physics: String theory and its Applications: From meV to the Planck Scale,* 3–46, 10 (2010).
4. Maldacena, J. The large-$N$ limit of superconformal field theories and supergravity. *Int. J. Theor. Phys.* **38**(4), 1113–1133 (1999).
5. Klebanov, I. R. & Polyakov, A. M. AdS dual of the critical O(N) vector model. *Phys. Lett. B* **550**(3–4), 213–219 (2002).
6. Strominger, A. Black hole entropy from near horizon microstates. *JHEP* **02**, 009 (1998).
7. Iqbal, N., Liu, H. & Mezei, M. Lectures on holographic non-Fermi liquids and quantum phase transitions. In *Theoretical Advanced Study Institute in Elementary Particle Physics: String theory and its Applications: From meV to the Planck Scale*, 707–816, 10 (2011).
8. Ryu, S. & Takayanagi, T. Holographic derivation of entanglement entropy from AdS/CFT. *Phys. Rev. Lett.* **96**, 181602 (2006).
9. Ryu, S. & Takayanagi, T. Aspects of holographic entanglement entropy. *JHEP* **08**, 045 (2006).
10. Faulkner, T., Leigh, R. G., Parrikar, O. & Wang, H. Modular hamiltonians for deformed half-spaces and the averaged null energy condition. *JHEP* **09**, 038 (2016).
11. Hartman, T., Kundu, S. & Tajdini, A. Averaged null energy condition from causality. *JHEP* **07**, 066 (2017).
12. Balakrishnan, S., Faulkner, T., Khandker, Z. U. & Wang, H. A general proof of the quantum null energy condition. *JHEP* **09**, 020 (2019).
13. Kraus, P. Lectures on black holes and the $AdS_3/CFT_2$ correspondence. *Lect. Notes Phys.* **755**, 193–247 (2008).
14. Banados, M. Notes on black holes and three-dimensional gravity. *AIP Conf. Proc.* **490**(1), 198–216 (1999).
15. Carlip, S. Conformal field theory, (2+1)-dimensional gravity, and the BTZ black hole. *Class. Quant. Grav.* **22**, R85–R124 (2005).
16. David Brown, J. & Henneaux, M. Central charges in the canonical realization of asymptotic symmetries: An example from three dimensional gravity. *Commun. Math. Phys.* **104**(2), 207–226 (1986).
17. Leutwyler, H. A (2+1)-dimensional model for the quantum theory of gravity. *Il Nuovo Cimento A (1971–1996)* **42**(1), 159–178 (1966).
18. Martinec, E. Soluble systems in quantum gravity. *Phys. Rev. D* **30**, 1198–1204 (1984).
19. Deser, S., Jackiw, R. & Hooft, G. Three-dimensional Einstein gravity: Dynamics of flat space. *Ann. Phys.* **152**(1), 220–235 (1984).
20. Maloney, A. & Witten, E. Quantum gravity partition functions in three dimensions. *JHEP* **02**, 029 (2010).
21. Liam Fitzpatrick, A., Kaplan, J. & Walters, M. T. Universality of long-distance AdS physics from the CFT bootstrap. *J. High Energy Phys.* **2014**(8), 1–65 (2014).
22. Hijano, E., Kraus, P., Perlmutter, E. & Snively, R. Semiclassical Virasoro blocks from $AdS_3$ gravity. *JHEP* **12**, 077 (2015).
23. Liam Fitzpatrick, A., Kaplan, J., Li, D. & Wang, J. On information loss in $AdS_3/CFT_2$. *JHEP* **05**, 109 (2016).
24. Kravchuk, P. Casimir recursion relations for general conformal blocks. *J. High Energy Phys.* **2018**(2), 1–82 (2018).
25. Zamolodchikov, A. B. Conformal symmetry in two dimensions: An explicit recurrence formula for the conformal partial wave amplitude. *Commun. Math. Phys.* **96**(3), 419–422 (1984).
26. Zamolodchikov, A. B. Conformal symmetry in two-dimensional space: Recursion representation of conformal block. *Theor. Math. Phys.* **73**(1), 4 (1988).
27. Zamolodchikov, A. B. & Fateev, V. A. Disorder fields in two-dimensional conformal quantum-field theory and N=2 extended supersymmetry. *Soviet J. Exp. Theor. Phys.* **63**(5), 913 (1986).
28. Harlow, D., Maltz, J. & Witten, E. Analytic Continuation of Liouville. *Theory. JHEP* **12**, 071 (2011).
29. de Boer, J., Castro, A., Hijano, E., Jottar, J. I. & Kraus, P. Higher spin entanglement and $W_N$ conformal blocks. *J. High Energy Phys.* **2015**(7), 1–49 (2015).
30. Hartman, T. Entanglement Entropy at Large Central Charge (2013).
31. Chen, B. & Wu, J.-q. Holographic entanglement entropy for a large class of states in 2D CFT. *J. High Energy Phys.* **2016**(9), 1–7 (2016).
32. Gerbershagen, M. Monodromy methods for torus conformal blocks and entanglement entropy at large central charge. *JHEP* **08**, 143 (2021).
33. Banerjee, P., Datta, S. & Sinha, R. Higher-point conformal blocks and entanglement entropy in heavy states. *J. High Energy Phys.* **2016**(5), 1–41 (2016).
34. Zamolodchikov, A. B. Two-dimensional conformal symmetry and critical four-spin correlation functions in the Ashkin-Teller model. *Sov. Phys. JETP* **63**, 1061–1066 (1986).
35. Beşken, M., Datta, S. & Kraus, P. Semi-classical Virasoro blocks: Proof of exponentiation. *J. High Energy Phys.* **2020**(1), 1–6 (2020).
36. Faulkner, T. & Wang, H. Probing beyond ETH at large $c$. *J. High Energy Phys.* **2018**(6), 1–41 (2018).
37. Liam Fitzpatrick, A., Kaplan, J. & Walters, M. T. Virasoro conformal blocks and thermality from classical background fields. *JHEP* **11**, 200 (2015).
38. Balasubramanian, V., Bernamonti, A., Craps, B., De Jonckheere, T. & Galli, F. Heavy-Heavy-Light-Light correlators in Liouville theory. *JHEP* **08**, 045 (2017).
39. Beccaria, M., Fachechi, A. & Macorini, G. Virasoro vacuum block at next-to-leading order in the heavy-light limit. *JHEP* **02**, 072 (2016).
40. Asplund, C. T., Bernamonti, A., Galli, F. & Hartman, T. Holographic Entanglement Entropy from 2d CFT: Heavy States and Local Quenches. *JHEP* **02**, 171 (2015).
41. Di Francesco, P., Mathieu, P. & Sénéchal, D. *Conformal Field Theory. Graduate Texts in Contemporary Physics* (Island Press, 1996).
42. Alkalaev, K. & Belavin, V. Monodromic vs geodesic computation of Virasoro classical conformal blocks. *Nucl. Phys. B* **904**, 367–385 (2016).

43. Alkalaev, K. Many-point classical conformal blocks and geodesic networks on the hyperbolic plane. *J. High Energy Phys.* **2016**, 70 (2016).
44. Ferrara, S., Grillo, A. F. & Gatto, R. Tensor representations of conformal algebra and conformally covariant operator product expansion. *Ann. Phys.* **76**(1), 161–188 (1973).
45. Rattazzi, R., Rychkov, V. S., Tonni, E. & Vichi, A. Bounding scalar operator dimensions in 4D CFT. *JHEP* **12**, 031 (2008).
46. Menotti, P. Classical conformal blocks. *Mod. Phys. Lett. A* **31**, 1650159 (2016).
47. Menotti, P. Torus classical conformal blocks. *Mod. Phys. Lett. A* **33**, 1850166 (2018).
48. Pavlov, M. Example of the 4-pt Non-vacuum $W_3$ Classical Block (2022).
49. Hadasz, L., Jaskólski, Z. & Suchanek, P. Elliptic recurrence representation of the $N=1$ Neveu-Schwarz blocks. *Nucl. Phys. B* **798**(3), 363–378 (2008).
50. Alkalaev, K. & Belavin, V. Large-$c$ superconformal torus blocks. *J. High Energy Phys.* **2018**(8), 1–21 (2018).

### Acknowledgements
The author would like to thank Prof. Huajia WANG for suggesting the idea, guiding, and improving the draft; the author would also like to thank Xiang LIU, Yang LEI, Jie-qiang WU, Jia TIAN, and Jiasheng LIU for useful discussions and comments on the draft. The author thanks the financial support from the UCAS Young Talent Nurture and Support Grant under Grant No. E0EG4302X2. The work is also supported by the National Key R&D Program of China under grant No. 2018YFA0404602.

### Author contributions
Y.H. wrote the main manuscript text. All authors reviewed the manuscript.

### Competing interests
The author declare no competing interests.

### Additional information
**Supplementary Information** The online version contains supplementary material available at https://doi.org/10.1038/s41598-022-16054-0.

**Correspondence** and requests for materials should be addressed to Y.H.

**Reprints and permissions information** is available at www.nature.com/reprints.

**Publisher's note** Springer Nature remains neutral with regard to jurisdictional claims in published maps and institutional affiliations.

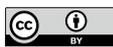